 \documentstyle[prl,aps,epsfig,preprint]{revtex}

\begin{document}

\draft

 \preprint{Faraday}

\title{
 Enhancement of Magneto-Optic Effects via Large Atomic Coherence
}

\author{
 V.~A.~Sautenkov$^{1,5}$,
 M.~D.~Lukin$^2$,
 C.~J.~Bednar$^{1,3}$,
 G.~R.~Welch$^1$,
 M.~Fleischhauer$^{1,4}$,
 V.~L.~Velichansky$^{1,5}$,
 and
 M.~O.~Scully$^{1,3}$
}

\address{%
 $^1$Department of Physics,
 Texas A\&M University,
 College Station, Texas 77843-4242,%
}
\address{%
 $^2$ ITAMP,
 Harvard-Smithsonian Center for Astrophysics,
 Cambridge, MA 02138,%
}
\address{%
 $^3$Max-Plank-Institut f\"ur Quantenoptik,
 Garching, D-85748, Germany,%
}
\address{%
 $^4$Sektion Physik,
 Universit\"at M\"unchen,
 D-80333 M\"unchen, Germany%
}
\address{%
 $^5$ Lebedev Institute of Physics,
 Moscow, Russia%
}

\date{\today}

\maketitle

\begin{abstract}

	We utilize the generation of large atomic coherence
to enhance the resonant nonlinear magneto-optic effect
by several orders of magnitude, thereby eliminating power
broadening and improving the fundamental signal-to-noise ratio.
A proof-of-principle experiment is carried out in a dense vapor
of Rb atoms.  Detailed numerical calculations are in good
agreement with the experimental results.  Applications such
as optical magnetometry or the search for violations of parity
and time reversal symmetry are feasible.

\end{abstract}


 \newpage

	Resonant magneto-optic effects such as the nonlinear
Faraday and Voigt effects~\cite{Gawlik74,Drake88} are important
tools in high-precision laser spectroscopy.  Applications to
both fundamental and applied physics include the search for
parity violations~\cite{CPT,Budker} and optical magnetometry.
In this Letter, we demonstrate that the large atomic coherence
associated with Electromagnetically Induced Transparency
(EIT)~\cite{Harris97,Gaeta} in optically thick samples can be
used to enhance nonlinear-Faraday signals by several orders of
magnitude while improving the fundamental signal-to-noise ratio.

	There exists a substantial
body of work on nonlinear magneto-optical
techniques~\cite{Gawlik74,Drake88,Budker,NLF,Budker-old,weis-vas,Kanorsky93,Kanorsky95},
which have been studied both in their own right and for
applications.  Such techniques can achieve high sensitivity
in systems with ground-state Zeeman sublevels due to the
narrow spectroscopic features associated with coherent
population trapping~\cite{Arimondo95}.  The ultimate
width of these resonances is determined by the lifetime
of ground-state Zeeman coherences, which can be made
very long by a number of methods (buffer gases and/or wall
coating~\cite{Budker,Kanorsky95,Meschede97} in vapor cells, or
atomic cooling or trapping techniques).  These resonances are
easily saturated, however, and power broadening deteriorates
the resolution even for very low light intensities.  For this
reason, earlier observations of nonlinear magneto-optic features
used small light intensities and optically thin samples,
which corresponds to a weak excitation of Zeeman coherences.
Recently, remarkable experiments of Budker and co-workers
demonstrated an excellent performance of the magento-optic
techniques in this regime:  very narrow magnetic resonances
were observed in a cell with a special paraffin coating~\cite{Budker}
(effective Zeeman relaxation rate $\gamma_0 \sim 2 \pi \cdot
1$Hz).

	This Letter shows that by increasing atomic density
and light power simultaneously the magneto-optic signal
can be enhanced substantially and the fundamental noise
(shot noise) can be greatly reduced.  A nearly maximal
Zeeman coherence generated under these conditions preserves
the transparency of the medium despite the fact that the
system operates with a density-length product that is many
times greater than that appropriate for 1/e absorption of a
weak field.  At the same time this medium is extraordinarily
dispersive~\cite{Lukin97,Scully92}, such that even very weak
magnetic fields lead to a large magneto-optic rotation.  This
effect is of the same nature as those resulting in ultra-low
group velocities~\cite{hau}.  Our experimental results
show a potential for several orders of magnitude improvement
over the conventional thin-medium--low-intensity approach.

	Typical measurements of the nonlinear Faraday effect
involve an ensemble of atoms with ground-state Zeeman sublevels
interacting with a linearly polarized laser beam.  In the
absence of a magnetic field, the two circularly-polarized
components generate a coherent superposition of the ground-state
Zeeman sublevels corresponding to a dark state.  A weak
magnetic field $B$ applied to such an atomic ensemble causes a
splitting of the sublevels and induces phase shifts $\phi_{\pm}$
which are different for right (RCP) and left (LCP) circularly
polarized light.  Hence, as linearly polarized light passes
through the medium, the direction of polarization changes by
an angle $\phi$ due to the differing changes in the phase of
the two circular components.

\begin{figure}[ht]
 \centerline{\epsfig{file=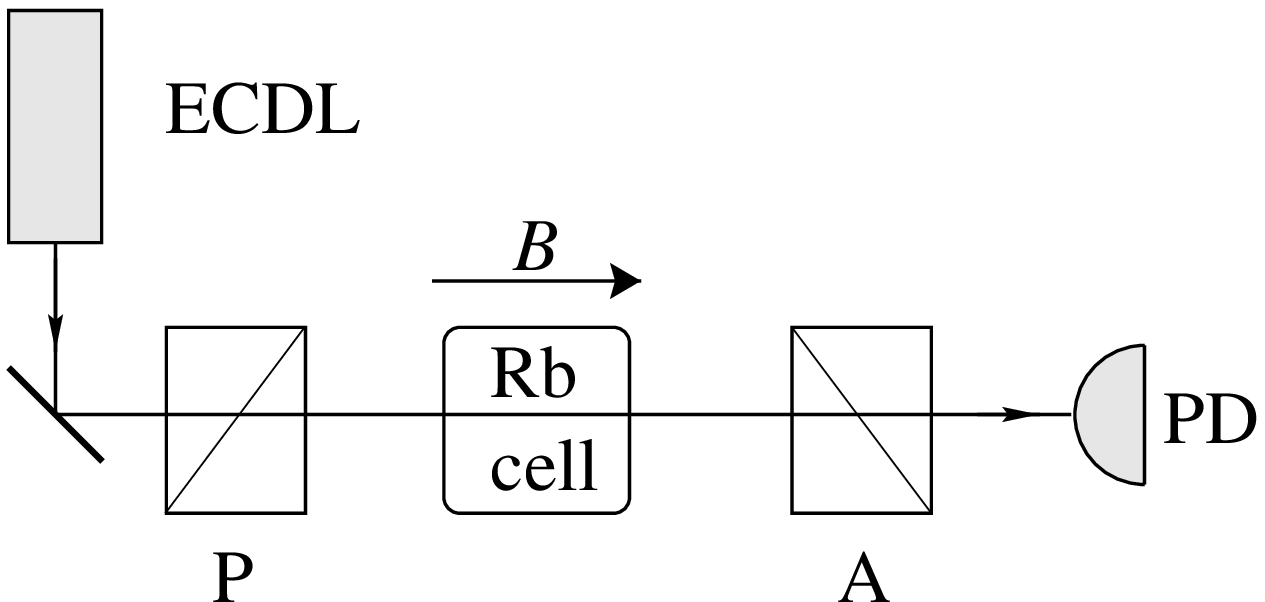,width=8cm}}
 \vspace*{2ex}
 \caption{
 	\label{exp.fig}
	Schematic of the experimental setup.  The laser beam
	passes through polarizer (P), Rb cell, and analyzer
	(A) and is detected by the photodiode (PD).
 }
\end{figure}

	In our experiment, shown schematically in
Fig.~\ref{exp.fig}, an external cavity diode laser (ECDL)
was tuned to the 795 nm $F=2 \rightarrow F'=1$ transition
of the $^{87}$Rb D$_1$ absorption line.  The laser beam was
collimated with a diameter of 2 mm and propagated through a
3 cm long magnetically shielded vapor cell placed between two
crossed polarizers.  The cell was filled with natural Rb and a
Ne buffer gas at a pressure of 3 Torr.  The laser power was 3
mW at the cell entrance.  The cell was heated to produce atomic
densities of $^{87}$Rb near $10^{12}$ cm$^{-3}$.  A longitudinal
magnetic field was created by a solenoid placed inside the
magnetic shields and was modulated at a rate of about 10 Hz.
The ground state relaxation rate was measured by decreasing
sufficiently the laser power, decreasing the density until
the absorption was low, and using RF-optical double resonance
techniques.  The measured value of $\gamma_0 \approx 2 \pi \cdot
5$ kHz (FWHM) is attributed to time-of-flight broadening as well
as to a residual inhomogeneous magnetic field.  The frequency
scale of the magnetic resonance was also determined by using
an RF-optical double resonance method.

\begin{figure}[ht]
 \centerline{\epsfig{file=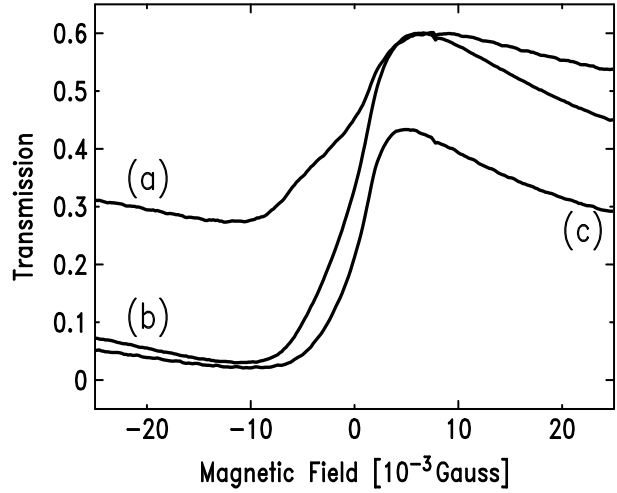,width=8.0cm}}
 \vspace*{2ex}
 \caption{
 	\label{data.fig}%
	Experimentally measured transmission through the system
	of Fig.~\ref{exp.fig} where the polarizer axes are
	$45^\circ$ apart.  The vertical scale is normalized
	such that unity corresponds to the transmission of the
	laser beam in the absence of the atomic cell and the
	polarizers.  Thus, zero magnetic field at low density
	gives a transmission through the system of 50\%.
	Curves (a-c) correspond to increasing atomic density:
	(a) $N = 3 \times 10^{11}$ cm$^{-3}$, (b) $N = 1 \times
	10^{12}$, (c) $N = 2 \times 10^{12}$.
 }
\end{figure}

	Figure~\ref{data.fig} shows the result of direct
measurement of the laser intensity at the photo-detector after
transmission through the system of two crossed polarizers
($\theta = 45$ degrees) and a vapor cell.  We emphasize
that no lock-in detection has been used for the data shown
in Fig.~\ref{data.fig} whereas in typical nonlinear Faraday
measurements sophisticated detection techniques are usually
required to obtain a reasonable signal-to-noise ratio.
We note that magneto-optical rotation angles increase with
optical density as does the slope $\partial \phi/ \partial B$
(curves a and b in Fig.~\ref{data.fig}).  The latter increase
is the essence of the method being described.  Under the
present conditions, rotation angles up to 0.7 radians have
been observed (curve b) with a good signal-to-noise ratio.
For very high densities the absorption becomes large and
the amplitude of the magneto-optic signal does not grow with
density any further (curve c).

	From our measurements of the rotation angles we obtain,
for the conditions outlined above:
\begin{eqnarray}
\partial \phi/ \partial B = 1.8 \times 10^{2} \mathrm{rad/G}~.
\end{eqnarray}
To put this result in perspective, we can estimate the
shot-noise limited sensitivity of this medium.  The fundamental
photon-counting error accumulated over a measurement time
$t_m$ scales inversely with the output intensity.  That is,
for a laser frequency $\nu$~\cite{Budker}
\begin{equation}
\label{shot-noise}
\Delta\phi_{err}\sim \sqrt{\hbar \nu/[P(L) t_m]}
\end{equation}
where $P(L)$ is the power transmitted through the cell.
Combining this with our measured rotation angles
implies a shot-noise limited sensitivity $B_{min}
= \Delta\phi_{err}/(\partial\phi/\partial B)$ of about
10$^{-10}$ G/$\sqrt{\rm Hz}$, which is comparable to the best
values estimated in e.g. Ref.~\cite{Budker}.  It is important
to note that this high sensitivity is achieved in our case
{\it despite more than three orders of magnitude difference}
of the ``natural'' width of the Zeeman coherence $\gamma_0$.
This demonstrates the very significant potential of the
present technique.

\begin{figure}[ht]
 \centerline{\epsfig{file=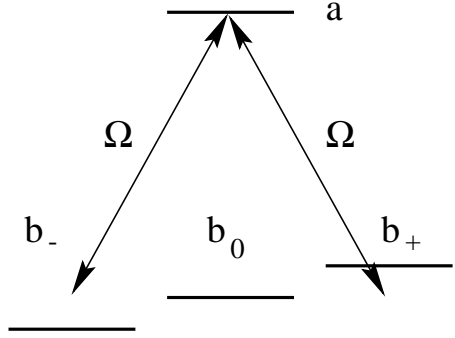,width=6cm}}
 \vspace*{2ex}
 \caption{
 	\label{simple.fig}
	A simplified, four-state model for observation of
	nonlinear Faraday effect.  $\Omega$ is the Rabi-frequency
	of ${\hat \sigma}_{\pm}$ components of an ${\hat
	x}$-polarized laser field. The magnetic field $B$
	shifts $m  = \pm 1$ levels by $\pm \delta$.
 }
\end{figure}

\begin{figure}[ht]
 \centerline{\epsfig{file=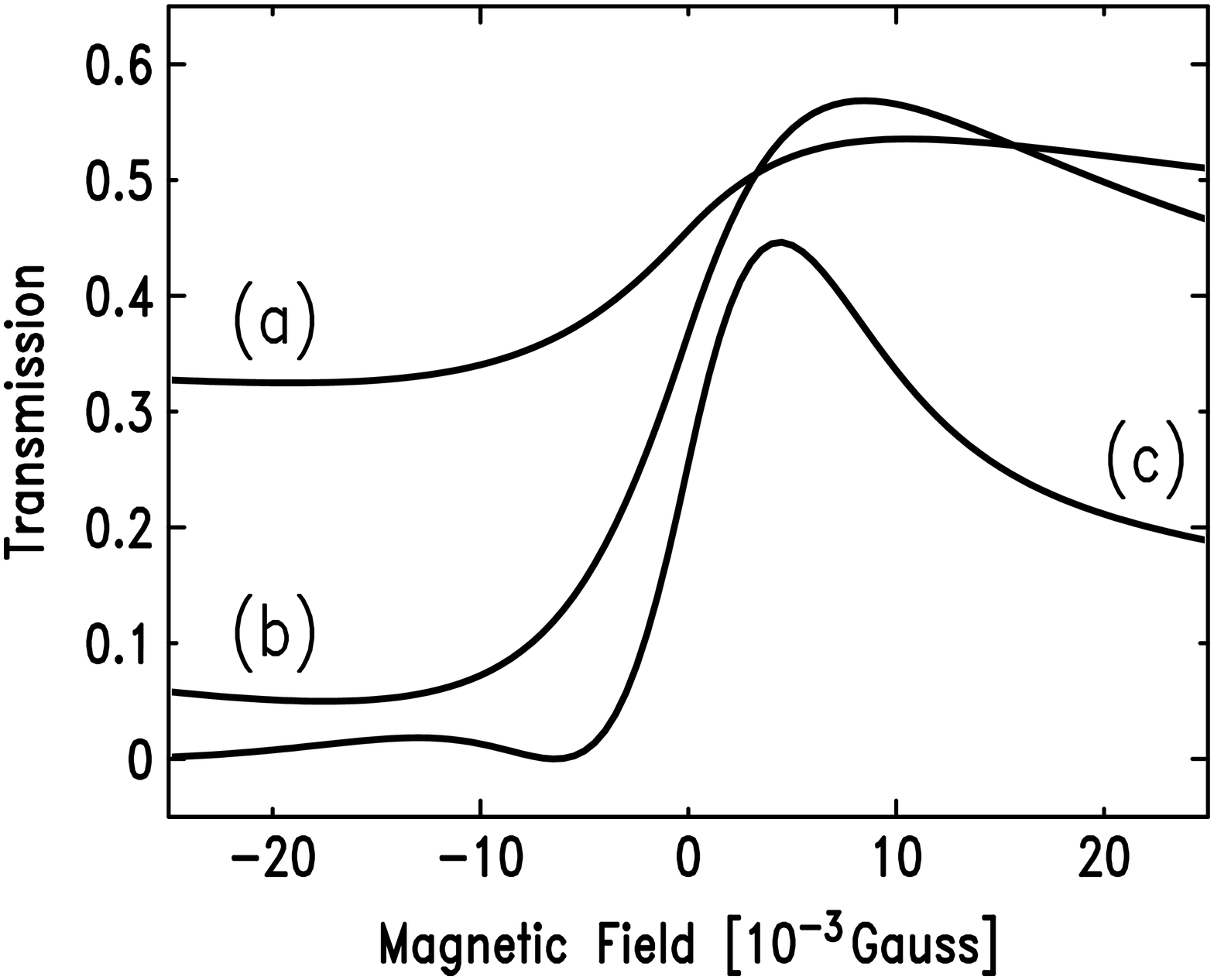,width=8.0cm}}
 \vspace*{2ex}
 \caption{
 	\label{calc.fig}%
	Results of numerical simulations with parameters
	corresponding to Fig.~\ref{data.fig}.
}
\end{figure}

	We now turn to a theoretical consideration of this
result.  As a simple model, let us consider the interaction of a
dense ensemble of atoms with ground-state angular momentum $F=1$
and an excited state $F=0$ as shown in Fig.~\ref{simple.fig}.
(Although the calculation presented in Fig.~\ref{calc.fig}
represents a simulation of realistic rubidium hyperfine
structure, this simple model, with well-chosen parameters,
represents the qualitative physics quite well.)  We consider a
strong laser tuned to exact resonance with the atomic transition
and disregard inhomogeneous broadening.  RCP and LCP intensities
are then attenuated equally according to:
\begin{eqnarray}
 \frac{1}{P} \frac{d P}{d z}
   &=&
       \kappa \gamma \phantom{\delta} \frac{
        \left[ 2 |\Omega|^2 \gamma_0 + \gamma (4 \delta^2 + \gamma_0^2) \right]
          \Delta \rho
       }{
        (2 |\Omega|^2 + \gamma \gamma_0 - 2 \delta^2)^2
        + \delta^2 (2 \gamma + \gamma_0)^2
       } \;, \label{a} \\
 \frac{d \phi_{\pm}}{d z}
   &=&  \pm \frac{\kappa \gamma \delta}{2}
       \frac{
        \left[ 4|\Omega|^2 - 4 \delta^2 - \gamma_0^2 \right] \Delta \rho
       }{
        (2 |\Omega|^2 + \gamma \gamma_0 - 2 \delta^2)^2
          + \delta^2 (2 \gamma+ \gamma_0)^2
       } \;, \label{re}
\end{eqnarray}
where $\Omega = \wp |E_{\pm}|/\hbar$ are the (equal) Rabi
frequencies of the field components ($P \propto |\Omega|^2$),
$\gamma_0$ and $\gamma$ are the relaxation rate of Zeeman
and optical coherences respectively, $\delta = g\mu_B
B/\hbar$ is the Zeeman level shift caused by an magnetic
field $B$ (g is a Land\'e factor), and $\kappa = 3 /(4 \pi)
{N \lambda^2} (\gamma_{a\rightarrow b}/ \gamma)$ is the weak
field absorption coefficient (inverse absorption length), and
$\gamma_{a\rightarrow b}$ is the natural width of the resonance.
The population difference between the ground-state Zeeman
sublevels and the upper state is $\Delta \rho$.  This quantity
is affected by optical pumping into the decoupled states
($b_0$ in Fig.~\ref{simple.fig}), and depends upon
cross-relaxation rates and applied magnetic field.  For a weak
magnetic field, $\Delta \rho \approx 1/3$.

	One recognizes from Eq.~(\ref{re}) that in the
case of optically thin media ($\kappa L \le 1$ where $L$
is the cell length) the phase shifts $\phi_\pm$ can be
approximated by dispersive Lorentzian functions of $\delta$,
with amplitude $\phi_{max} = \kappa L \Omega^2/(2 \Omega^2
+ \gamma\gamma_0) \Delta \rho$ and width $\delta_0 =
\gamma_0/2 + \Omega^2/\gamma$.  The former is typically
rather small (on the order of mrad in the experiments of
Ref.~\cite{Gawlik74,Drake88,Budker,Budker-old,weis-vas,Kanorsky93,Kanorsky95})
while the latter saturates when $|\Omega|^2$ exceeds
the product $\gamma\gamma_0/2$, which corresponds
to the usual power-broadening of the magneto-optic
resonance~\cite{Budker,weis-vas}.

	It is important to emphasize here that a principal
difference between regimes involving low and high driving
power lies in the degree of Zeeman coherence excited by the
optical field:   
\begin{equation}
  \rho_{b_{-} b_{+}}
      = \frac{
         2 | \Omega |^{2} \Delta \rho
        }{
         2 | \Omega |^{2}
          + \gamma \gamma_{0} - 2 \delta^{2}
          + i \delta ( 2 \gamma + \gamma_{0} )
        }
  \label{coh}
\end{equation}
Large coherence corresponds to a large population difference
between symmetric (i.e.  ``bright'') and antisymmetric
(i.e. ``dark'') superpositions of Zeeman sublevels. In the
low-power regime this difference is small corresponding to a
small coherence. In a regime where the width of the resonance
is determined by saturation, a very large (nearly maximal)
Zeeman coherence is generated, as per Eq.(\ref{coh}).

	We will now show that in a medium with large
Zeeman coherence the magneto-optic signal is maximized if
a large density-length product is chosen.  In the case of
a strong optical field ($|\Omega|^2 \gg \gamma_0\gamma$)
and weak magnetic fields $|\delta| < |\Omega|^2/\gamma,
\sqrt{\gamma_0/\gamma}|\Omega|$, integration of Eqs.~(\ref{a})
and (\ref{re}) yields for the transmitted power and the rotation
angle $\phi = ( \phi_+ - \phi_- ) / 2$
\begin{eqnarray}
P(L)
   &=& (1 - \alpha_0 L) \, P(0) \\
 \phi(L)
   &=& \frac{\delta}{2 \gamma_0} \,
        \ln \left[
           \frac{1}{1 - \alpha_0 L}
        \right]
 \label{signal}
\end{eqnarray}
where $\alpha_0 = \Delta \rho \; \kappa \gamma\gamma_0/
2 |\Omega_0|^2$, and $\Omega_0$ corresponds to the input field.

	Note that in the case of a strong input field
and an optically thin medium $\alpha_0 L \ll 1$.  However
Eq.~(\ref{signal}) shows that maximal rotation is achieved with
a large density-length product $\alpha_0 L$.  Clearly $\alpha_0
L$ cannot be too close to unity, since then no light would
be transmitted.  Using Eq.~(\ref{shot-noise}) one finds the
optimum value $(\alpha_0 L)_{opt}=1-{\rm e}^{-2}$, corresponding
to a density-length product
\begin{equation}
 \Delta\rho \kappa L \vert_{opt}
   = (\alpha_0 L)_{opt} \,
       \frac{|\Omega_0|^2}{\gamma \gamma_0}
   \gg 1 ~.
\end{equation}
In this case the total accumulated rotation angle is quite large and
the slope of its dependence upon $B$ is maximal:
\begin{equation}
 \partial \phi_{opt} / \partial B
   = g\mu_B / (\hbar\gamma_0)
\end{equation}
which is in strikingly good agreement with our measured value.
The significance of this result can be understood by noting that
in a shot-noise limited measurement, the minimum detectable
rotation $\phi_{err}$ given in Eq.~(\ref{shot-noise}) is
inversely proportional to the square root of the laser power.
Working at high power, therefore, has a clear advantage, since
the fundamental shot noise error is reduced even though the
signal is large.

	To make a more realistic comparison of theory and
experiment, we have carried out detailed calculations in
which coupled density matrix and Maxwell equations including
propagation through the medium and Doppler broadening have been
solved numerically for the two components of the optical field.
The calculation takes into account a 16-state atomic system
with energy levels and coupling coefficients corresponding to
those of the Rb $D_1$ line.  The results of these calculations
are shown in Fig.~\ref{calc.fig} and are in good agreement
with the experimental results.  In particular, we note that
our calculations predict the maximal rotation angle, (which is
apparently limited by the optical pumping into the F=1 $S_{1/2}$
hyperfine manifold) as well as the slope of the resonance curve.

	It is important to comment at this point on possible
limitations for the extension of the present technique into
the domain of narrow resonances.  For instance, in the case of
a long-lived ground state coherence, spin-exchange collisions
can become a limiting factor for the Zeeman relaxation rate.
In the case of Rb, this is a few tens of Hz at densities
corresponding to the present operating conditions.  We note
however that it is possible to operate at lower densities
by increasing the optical path length (e.g.\ by utilizing
an optical cavity).  Likewise, the role of the light shifts
due to off-resonant coupling to e.g.\ F=1 $S_{1/2}$ and
F=2 $P_{1/2}$ hyperfine manifolds needs to be clarified.
Although the noise due to classical intensity fluctuations
of the circular components of the optical field is obviously
canceled in a measurement of the polarization rotation, there
might exist additional quantum contributions that add noise.
We note however that even if such contributions are present,
it is likely that they can be suppressed by tuning the laser
to the point of minimum light shifts.

	For these reasons, we believe that the combination
of the present approach with buffer gas or wall-coating
techniques is likely to improve substantially the sensitivity
of nonlinear magneto-optical measurements.  Therefore, we
anticipate that this method will be of interest for sensitive
optical magnetometry as well as for setting new, lower bounds
in test for the violation of parity and time-reversal invariance
\cite{CPT,Budker}.

	The authors warmly thank Leo Hollberg, Alexander Zibrov,
and Michael Kash for useful discussions and Tamara Zibrova for
valuable assistance.  We gratefully acknowledge the support of
the Office of Naval Research, the National Science Foundation,
the Welch Foundation, and the Air Force Research Laboratory.



 \newpage

\end{document}